\documentclass[11pt,twocolumn]{IEEEtran}
\ifCLASSINFOpdf
   \usepackage[pdftex]{graphicx}
\else
\fi
\usepackage[table,xcdraw]{xcolor}
\usepackage{textcase}
\usepackage[tablename=TABLE]{caption}

\usepackage{amsmath}
\usepackage{amsfonts}

\begin{document}
\title{A Survey on Poisoning Attacks Against Supervised Machine Learning}

\author{Wenjun Qiu\\
Electrical and Computer Engineering\\University of Toronto\\
wenjun.qiu@mail.utoronto.ca}


%Telephone: (800) 555--1212, Fax: (888) 555--1212}


\maketitle

\begin{abstract}
With the rise of artificial intelligence and machine learning in modern computing, one of the major concerns regarding such techniques is to provide privacy and security against adversaries. We present this survey paper to cover the most representative papers in poisoning attacks against supervised machine learning models. We first provide a taxonomy to categorize existing studies, and then present detailed summaries for selected papers. We summarize and compare the methodology and limitations of existing literature. We conclude this paper with potential improvements and future directions to further exploit and prevent poisoning attacks on supervised models. We propose several unanswered research questions to encourage and inspire researchers for future work.
\end{abstract}


% \fi
\IEEEpeerreviewmaketitle

\section{Introduction}
Machine learning has become one of the most popular techniques for a wide variety of areas and applications, as it provides models the ability to automatically learn and improve from experience while no explicit algorithm is required to be programmed. With such contributions and impact on academia, industry and society, it is crucial to draw our attention to an emerging concern: the privacy and security of machine learning. Specifically, many machine learning models suffer from vulnerabilities that could be exploited by experienced attackers. These models can be manipulated in various ways, resulting in a negative impact on the model performance, leakage in private data, or misbehaviours of the machine learning systems. It is important to recognize such vulnerabilities and study their countermeasures to prevent model corruptions. In this paper, we focus on the setting of poisoning attacks, where adversaries inject a small amount of poisoned samples during the training process. Such poisoned samples are generated by the attackers to maximize their incentives and manipulate the results and generated models. Examples of poisoning attacks on spam filters~\cite{miller2018mixture}, fraud detection~\cite{fraud_detect},  malware detection~\cite{wang2017adversary}, digit recognizer~\cite{yang2017generative} have demonstrated their successes in practice.

\begin{figure}
    \centering
    \includegraphics[scale=0.61]
    {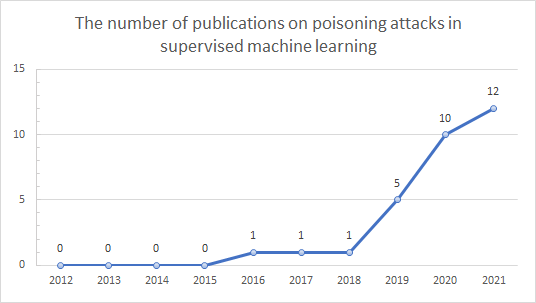}
    \caption{The popularity of poisoning attacks against machine learning models as a topic, approximated by the number of papers published each year with the term ``poisoning attack'' and ``supervised learning'' in their titles and abstracts. The data are obtained from Google Scholar.}
    \label{fig:overview_trend}
\end{figure}

\begin{table*}[] 
\centering
\begin{tabular}{|l|l|l|l|l|l|l|l|l|l|}
\hline
\multicolumn{1}{|c|}{\textbf{Keywords/Year}}                                                 & \textbf{2013} & \textbf{2014} & \textbf{2015} & \textbf{2016} & \textbf{2017} & \textbf{2018} & \textbf{2019} & \textbf{2020} & \textbf{2021} \\ \hline
\textit{``poison attack machine learning''} - title and abstract                  & 0             & 1             & 0             & 0             & 2             & 9             & 7             & 16            & 22            \\ \hline
\rowcolor[HTML]{CBCEFB} 
\textit{``poisoning   attack supervised learning''} - title and   abstract            & 0             & 0             & 0             & 1             & 1             & 1             & 5             & 10            & 12            \\ \hline
\textit{``poisoning attack supervised learning''} - full data               & 3789         & 2836         & 2265         & 1293         & 1009         & 3489         & 3031         & 2263         & 6511         \\ \hline
\end{tabular}
\caption{A summary of different topics in poisoning attacks on machine learning models, approximated by the number of papers published each year with different keywords. The data are obtained from Google Scholar. The highlighted line is graphed as Figure~\ref{fig:overview_trend} shows a line chart for the highlighted line.}
\label{tab:popularity}
\end{table*}

Poisoning attacks on machine learning models have been discussed in many existing papers, mostly being identified as a potential security issue. However, few researcher actually focus their study on this specific topic. As shown in Table~\ref{tab:popularity}, there are thousands of publications mentioning ``poisoning attack'' and ``supervised learning'' (number obtained by searching within papers' full metadata). In contrast, the number of papers include the same keywords in their titles and abstracts is no more than $30$. More importantly, related work only started to appear since 2014 (with the keyword ``machine learning'') and 2016 (with the keyword ``supervised learning''). We highlight the fact that the study of poisoning attacks on machine learning models is a relatively new field with a great future potential. 

To better present the popularity trend of this research direction, we record the number of related work on poisoning attacks against supervised machine learning models. Figure~\ref{fig:overview_trend} summarizes the trend of popularity by counting the total number of publications with the term ``poisoning attack'' and ``supervised learning'' appear in the paper title and abstract, recording since 2012. We collected the data from Google Scholar. As the Figure indicated, we observe a continuous rising trend in the popularity of poisoning attacks on supervised machine learning, despite having a late start only since 2016.  

In this paper, we present a list of studies on poisoning attacks and defenses for machine learning models. We include a summary of the popular papers and present our views and comments on these designs. We analyze and compare these attacks, and highlight the key insights on poisoning attacks against machine learning systems. We hope that this paper serves as a foundation for future research and potential improvements on the security and privacy of machine learning.

\section{Poisoning Attack Overview}
In this section, we will first describe some common threat models for poisoning attacks on machine learning models (Section~\ref{subsec:threat_model}). Then, we briefly summarize the steps to develop an effective ML attack (Section~\ref{subsec:ml_poison}).

\subsection{Threat Model} \label{subsec:threat_model}

\begin{figure*} [ht!]
    \centering
    \includegraphics[scale=0.39]
    {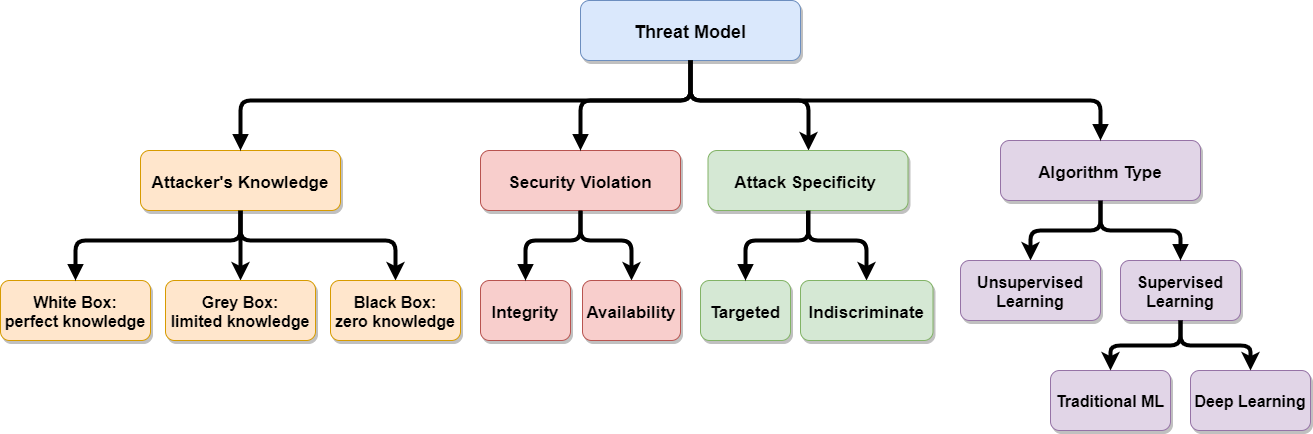}
    \caption{Taxonomy of Threat Models for Poisoning Attacks in Machine Learning}
    \label{fig:taxonomy}
\end{figure*}

In this section, we describe various threat models, covering the information on attacker's knowledge, security violation, and algorithm type. Figure~\ref{fig:taxonomy} summarizes the taxonomy we use to categorize existing literature on ML poisoning attacks.

\noindent\textbf{Attacker's Knowledge} indicates how much information the attackers have. For example, a whitebox attack assumes adversaries have the access to both the training algorithm including model parameters and the labelled samples, that is, perfect knowledge; whereas a greybox setting has either and a blackbox attack has none, in other words, limited knowledge for the former, and zero knowledge for the latter. 

\noindent\textbf{Security Violation} describes the end goal of an attacker. In poisoning attacks, they can either aim to violate the integrity or availability of a model, with the former being an evasion of the model's predictions, and the later being a compromise the normal system functionalities available to users. In other words, an integrity violation could introduce poisoning samples as a backdoor without affecting the normal ML inference process; an availability
violation, however, will disrupt the entire system. As examples, a backdoor for specific keyword in email exploits the model integrity, whereas a data injection attack to stop model from learning features and patterns targets the model availability. Other types of attacks may also target a privacy violation, however this is not in the scope of this paper. 

\noindent\textbf{Attack Specificity} highlights the results of an attack, ranging from a targeted attack to an indiscriminate attack. The former aims to produce mis-predictions from the original labels to a specifically targeted category, whereas the latter deviate the original labels to any mis-predicted categories without specification. As examples, a targeted attack in spam filtering could be passing spam on purpose or showing more advertisements, whereas an indiscriminate attack could simply aim to decrease the model accuracy. 

\noindent\textbf{Algorithm Type} indicates what type of machine learning algorithm the targeted model is based on. In general, machine learning studies can be categorized as supervised and unsupervised learning. In this survey paper, we focus on supervised machine learning only. Specifically, this paper covers poisoning attacks on traditional statistical models such as logistic regression (LR) and Support Vector Machines (SVM), and deep neural networks, including Convolutional Neural Network (CNN), Recurrent Neural Network (RNN) and so on. Note that we do not have any restriction on the model application. For example, machine learning models for vision image classification and natural language processing (NLP) could both be the victim of an ML poisoning attack. There are also studies on poisoning attacks and defenses for unsupervised learning models such as PCA \cite{2009_antidote}, however this is out of the scope of our paper.

\subsection{Machine Learning Poisoning} \label{subsec:ml_poison}

\begin{figure} [ht!]
    \centering
    \includegraphics[scale=0.5]
    {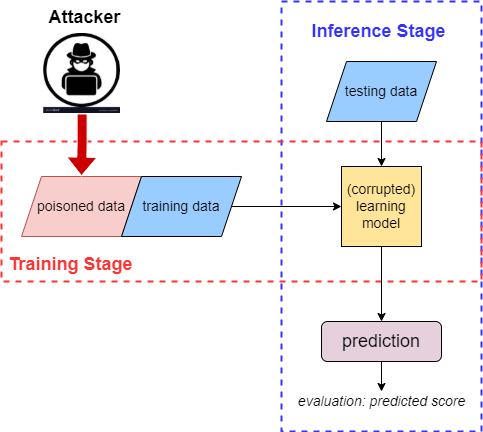}
    \caption{A generalized illustration for poisoning attacks on machine learning models}
    \label{fig:poison}
\end{figure} 

For the focus of this survey paper, we study existing literature on poisoning attacks with limited attacker's capability - training samples only. A general definition of poisoning attack is the adversarial action that involves polluting a machine learning model's training data, in contrast to attacks targeting test data. The high level goal of such attacks is to maximize the generalization error of the machine learning models, negatively affect the training progress and possibly mislead the predictions generated by the models. As we focus on supervised learning in this paper, we assume all samples have the potential to be correctly labeled. For the following sections, we denote the input space as $\mathbb{X} = \{x_0, x_1, ..., x_i, ..., x_n\}$, and the output space to be $\mathbb{Y} = \{y_0, y_1, ..., y_i, ..., y_n\}$ accordingly, where the prediction $y_i$ is generated by a utility function $f$ using the input sample $x_i$. In other words, $y_i = f(x_i)$. Following the same notation, we use $\hat{x_i}$ and $\hat{y_i}$ to represent a potential malicious system. The goal of an attacker is to maximize the negative impact introduced by the adversarial training points, for instance, to maximize $\left|f(\hat{x_i})-y_i\right|$. Figure~\ref{fig:poison} illustrates how a generalized version of poisoning attack works: an attacker gains control of partial data contribution to inject poisoning samples into the training data set at the training stage of a machine learning model. The ultimate questions for the research topic are:

\begin{enumerate}
  \item What is the goal of our poisoning attack? 
  \item How do we formalize this goal into a solvable mathematical problem? 
  \item How to craft poisoning samples using the derived mathematical formalization?  
\end{enumerate}
    
Essentially, the first step is to identify a target for attackers to exploit. For instance, one could focus on an indiscriminate setting to significantly decrease the model accuracy, or instead try to perturb training samples to cause a specific target test sample to be misclassified. This leads to the second research question, where researchers are required to formalize the identified goal. Many existing studies use a loss function as the reference to their attack, for example, Mean Squared Error (MSE). That is, because the goal is to minimize the training progress in terms of accuracy, attackers could use MSE as a metric to evaluate how effective the attack strategy and poisoning samples are. The last step is the creation of poisoning samples. In many cases, researchers view this question as an optimization problem. By solving a derived objective, they are able to craft the most influential poisoning samples with respect to the loss function. In other words, such poisoning samples are able to significantly increase the model loss, which is inversely proportional to the model performance. 

Due to various threat models and methodology, there are many different methods to produce such poisoning data. In the following paper (Section~\ref{sec:poison_ML}), we will further describe how existing work approach this research problem.

\section{Poisoning Attacks on Machine Learning} \label{sec:poison_ML}

In this section, we present a list of poisoning attacks with various threat models. We categorized these studies based on the attack methodology, as they exploit different vulnerabilities, even in similar machine learning models. These attacks target different security violations (integrity and/or availability) and attack specificity (targeted and/or indiscriminate). Table~\ref{table:compare_table} presents a comparative analysis of some representative papers on poisoning attacks against machine learning. In the following sections, we step forward and present more details on selected attack models, grouped by various attack methodology. We define \textit{attack methodology} as the underlying approaches to achieve attacker's adversarial goals. Specifically, we summarize these approaches into three categories:   
\begin{enumerate}
  \item \textbf{Labeling Flipping Attacks}, where an attacker creates poisoning samples by changing the label of selected training samples.   
  \item \textbf{Watermarking Attacks}, where an attacker creates poisoning samples by adding perturbations to training samples instead of labels, by merging and masking a target sample onto the training samples.  
  \item \textbf{Clean-Label Attacks}, where an attacker creates poisoning samples with adversarial perturbations that are ``invisible'' to human experts but harmful to the machine learning models. That is, the poisoning sample appear to be a benign training sample while negatively impacting the learning progress. 
\end{enumerate}

One of the common methodology to create poisoning samples is to solve a bi-level optimization problem, which is characterized by two objective functions. A general formulation of the bi-level optimization problem can be summarized as follows: 

\[\; \;\;\; \; \; \; \;\;\;\;\underset{x\in \mathbb{X}, y\in \mathbb{Y}}{arg min} \; F(x,y), 
\]
\[\mathbf{s.t.}\;  \; \; \; \; \underset{z\in \mathbb{Y}}{arg min} \; G(x,z) 
\]

\noindent where $F$ represents the upper-level objective function and $G$ represents the lower-level objective function; $x$ represents the upper-level decision vector and $y$ represents the lower-level decision vector. The lower-level objective function may be used to represent a constraint on the upper-level function. Note that the $argmin$ could be changed to $argmax$ trivially, as maximizing some objective function is equivalent to minimize its negative. 

Table~\ref{table:compare_table} provides a summary of the related work covered in this survey paper. The table outlines the attack methods, which include labeling flipping, watermarking, and clean-label attacks; machine learning model types, which include traditional supervised models and deep neural networks; attacker's violation, which includes integrity and availability; attack type, which can be targeted, indiscriminate, or both; and last but not least, attacker's knowledge, which includes white-box, grey-box and black-box. Note that due to the nature of poisoning attacks, it is necessary for the attackers to at least have some access to either the model or the data sets. As we can observe from the table, there is no attacks solely working under the black-box setting. Most black-box scenarios requires a surrogate models that mimic the target machine learning model. 

\begin{table*}[]
\centering
\begin{tabular}{|l|l|l|l|l|l|}
\hline
                                                                        & \textbf{Attack Method} & \textbf{ML Model}                                                                                                        & \textbf{\begin{tabular}[c]{@{}l@{}}Attacker’s Goal\\ (Violation)\end{tabular}} & \textbf{Attack Type}                                                & \textbf{\begin{tabular}[c]{@{}l@{}}Attacker’s \\ Knowledge\end{tabular}} \\ \hline
Barreno et al.~\cite{ref2}                        & Labeling flipping      & SpamBayes                                                                                                                & \begin{tabular}[c]{@{}l@{}}Integrity, \\ Availability\end{tabular}             & \begin{tabular}[c]{@{}l@{}}Targeted, \\ Indiscriminate\end{tabular} & White box                                                                \\ \hline
Paudice et al.~\cite{ref3}                        & Labeling flipping      & \begin{tabular}[c]{@{}l@{}}Linear classifier \\ (Hinge loss)\end{tabular}                                                & Integrity                                                                      & Indiscriminate                                                      & White box                                                                \\ \hline
Suciu et al.~\cite{ref5}                          & Watermark              & \begin{tabular}[c]{@{}l@{}}Customized Neural Networks:\\ (convolutional layers + \\ fully connected layers)\end{tabular} & Integrity                                                                      & Targeted                                                            & \begin{tabular}[c]{@{}l@{}}White box, \\ Grey box\end{tabular}           \\ \hline
Shafahi et al.~\cite{poison_frog}                & Watermark              & \begin{tabular}[c]{@{}l@{}}CNN: \\ InceptionV3, \\ AlexNet\end{tabular}                                                  & Integrity                                                                      & Targeted                                                            & \begin{tabular}[c]{@{}l@{}}White box, \\ Grey box\end{tabular}           \\ \hline
Biggio et al.~\cite{ref7}                         & Clean-label            & SVM                                                                                                                      & Integrity                                                                      & Indiscriminate                                                      & White box                                                                \\ \hline
Jagielski et al.~\cite{jagielski2021manipulating} & Clean-label            & LR                                                                                                                       & Integrity                                                                      & Indiscriminate                                                      & White box                                                                \\ \hline
Koh et al.~\cite{koh2020understanding}            & Clean-label            & \begin{tabular}[c]{@{}l@{}}SVM+RBF,\\ CNN (Inception V3)\end{tabular}                                                    & Integrity                                                                      & Indiscriminate                                                      & \begin{tabular}[c]{@{}l@{}}White box, \\ Black-box\end{tabular}          \\ \hline
Yang et al.~\cite{yang2017generative}             & Clean-label            & GAN + DNN                                                                                                                & Integrity                                                                      & Indiscriminate                                                      & White box                                                                \\ \hline
Dai et al.~\cite{dai2019backdoor}                 & Watermark              & LSTM                                                                                                                     & Integrity                                                                      & Targeted                                                            & Grey-box                                                                 \\ \hline
Shumailov et al.~\cite{reordering}                & Clean-label            & \begin{tabular}[c]{@{}l@{}}DNN: \\ ResNet, \\ LeNet, \\ MobileNet\end{tabular}                                           & Availability                                                                   & Indiscriminate                                                      & \begin{tabular}[c]{@{}l@{}}White box, \\ Black box\end{tabular}          \\ \hline
Shumailov et al.~\cite{sponge}                    & Clean-label            & \begin{tabular}[c]{@{}l@{}}DNN: \\ RoBERTa, \\ ResNet, \\ MobileNet\end{tabular}                                         & Availability                                                                   & Indiscriminate                                                      & \begin{tabular}[c]{@{}l@{}}White box, \\ Black box\end{tabular}          \\ \hline
\end{tabular}
\caption{A comparative analysis and summary of existing works on poisoning attacks against supervised machine learning models}
\label{table:compare_table}
\end{table*}

\subsection{Labeling Flipping Attacks}
As mentioned, labeling flipping attacks involve changing the labels of selected training samples. Despite being a possibly overly strong threat model, it is shown to be effective in practical machine learning~\cite{ref1}. As one of the earliest related work, Barreno et al.~\cite{ref2} introduce labeling flipping attacks using heuristic based modifications and demonstrate the ability to increase both false positives and false negatives of SpamBayes, a Bayesian anti-spam classifier to differentiate spam and ham emails. They craft poisoning samples by creating attack emails with a spam label, but adding benign tokens and phrases into the email texts. Their attacks can be either targeted or indiscriminate, depending on the desired outcomes. As such noises are introduced to the learning model, it is reasonable to result in a degraded performance. However, the attacks proposed in this paper are mostly hypothetical, and the heuristic based approach has limited generalization and transferability: does the attack work on other text classification tasks? Will the crafted poisoning samples also work on spam detectors trained on different machine learning models? Are there ways to extract the adversarial features from the crafted poisoning samples? These questions are now hot topics in machine learning interpretability, model privacy and machine unlearning. Although the paper did not provide answers to these questions, the proposed attacks serve as a foundation of poisoning attacks in machine learning, and prompt readers to think of potential improvements in the attacks. 

A more recent work by Paudice et al.~\cite{ref3} consider the problem of learning a binary linear classifier and create poisoning samples to maximize the loss function evaluated on a separate validation dataset. The output specifies the samples whose labels should be flipped. The attack is formalized as a bi-level optimization problem, which is intractable as it requires an exhaustive search amongst all possible combination of sample subsets. The authors propose a greedy algorithm to select samples to be flipped based on their impact on the validation objective function, and remove the selected samples for future search to reduce computation cost. Because they focus on maximizing the overall loss, the attack is indiscriminate, although in their experimental set up with binary classifiers using a linear model with hinge loss, the two attack specificity are essentially the same. Despite being an effective attack, the paper does not provide any controls on either the number of poisoning samples nor the selection of poisoned samples. An important consideration could be whether there should be a regularization term in the optimization function to minimize the number of poisoning samples needed to perform a successful attack. Another possible improvement might be to selectively poison samples only when the label flipping is not so obvious, therefore bypassing the defense mechanism. 

\subsection{Watermarking Attacks}

Another attacking methodology named watermarking is to perturb selected training samples instead of their labels. For this type of attacks, target samples are superimposed onto some training samples. The most significant advantage of watermarking attacks is that it does not require attackers to control the labels of training samples. It is also targeted because the poisoning samples are crafted based on a target instance as the reference. Moreover, because it targets on specific instance instead of negatively affecting the entire learning system, it is difficult to detect such attacks if using traditional defense mechanisms such as the Reject On Negative Impact (RONI)~\cite{ref2}, which evaluates the empirical effect of training samples and eliminates potential poisoning data that significantly degrade the model accuracy.

Shafahi et al.~\cite{poison_frog} design a feature collision attack in which they craft a poisoning sample that is closed to the target instance in the feature space (that is, the output prediction of the poisoning sample is similar to that of the target instance) while minimizing the distance between this poisoning sample and its original form, which is a given training sample, in the input space (that is, the poisoning sample should appear to be similar to a benign training sample). However, this approach fall short in cases where the target neural networks are deep with multiple trainable layers. The solution to this challenge is to extract a low-opacity watermark from the target instance and adding it to the corresponding poisoning sample, therefore prevents the separation of the two types of instance. A similar related work by Suciu et al.~\cite{ref5} uses the same watermarking technique but requires a larger amount of poisoning samples to achieve the same affect. Their experiments are performed on a customized neural network (NN) with multiple convolutional layers and fully connected layers.  

While having the same attack motivation, Dai et al.~\cite{dai2019backdoor} shift to more complex learning models: a deep neuron network. The authors propose a poisoning attack against an LSTM-based system for textual classification. They assume a grey-box setting where attackers have access to partial training samples but not the model type or parameters. They also injected adversarial samples as backdoor triggers, yet the triggers include obvious patterns that could be easily recognized by either human examiners. 

Despite being an excellent attack for backdoor poisoning, watermarking seems problematic when it comes to specific defense mechanisms~\cite{ref6}: as the opacity of watermark increases, it becomes easier for human experts to detect such attacks. It is also possible to develop automated detection systems to examine incoming training samples and determine whether the additional perturbations include influential features from some target instance. Controlling the extent of watermarking is the crucial key to perform a successful attack, which seems to require careful manual inspection rather than depending on simple distance measurements to guarantee the similarity between a base training sample and the crafted poisoning sample.

\subsection{Clean-Label Attacks}
Many existing studies demonstrate the effectiveness of the above two attacks. However, one of the most important shortcomings of these attacks is that they could be easily detected by human or automated supervision. On the other hand, a clean-label attack do not require control over the labels, while maintaining the original semantic of the training samples. In other words, even under the careful examination by human experts, poisoning samples that are ``clean'' are difficult to be detected, because they appear to have the correct labels based on human interpretations. However, such samples are harmful to the target learning model, in a way that the ``invisible'' noises perturb the model features and negatively impact the model weights to be shifted to the undesired decision boundary. 

The advantages of such attacks are obvious: it is very difficult to detect, and may open the door for successful attacks despite having no internal access to the data collection and labeling process. For example, attackers may create poisoning samples and upload them to openly available online platforms. Large models that require a massive amount of training data may use web scraping technique to collect samples, which may accidentally include the poisoning samples attackers uploaded beforehand. Another advantage is that such clean-label attacks are applicable to many machine learning models, including traditional statistical models as well as deep neural networks. It provides more potentials on attack generalization and transferability than labeling flipping and watermarking attacks. It requires less controls over the model data and training samples, opening up the opportunity for grey box and black box attacks.   

The aforementioned advantages of clean-label attacks encourage researchers to further study this topic. There are many existing work on ML poisoning attacks that fall into this category. To better illustrate their characteristics, we categorize them according to their target learning models, as shown in the following paragraphs:

\noindent\textbf{Linear Regression (LR)} is one of the simplest yet effective supervised learning models in many prediction tasks. Jagielski et al.~\cite{jagielski2021manipulating} propose a poisoning attack against LR model by using a gradient-based optimization framework to maximize the negative impact of training samples on a particular learning model. Specifically, they use gradient ascent to iteratively update training points and stop at convergence where they obtain a poisoned sample. The resulting poisoned samples are crafted to maximize the regularized loss function. Revisiting the bi-level optimization problem described above, this work falls into the same category: the upper-level ojbective function is to maximize the loss, whereas the lower-level objective function is the regular retrain process of any LR algorithm while including the poisoned samples crafted from the previous iteration. 

The paper also shows that this proposed attack is able to bypass existing defenses due to their common shortcoming: they only try to detect poisoned data by identifying outliers. The paper experimented on Ridge regression and LASSO, using three real-life datasets. The experiments cover both white-box and black-box scenarios, with the latter shown to be less efficient. One of the limitations of this paper is that the paper only covers experiments on two LR models. Further evaluation on, for example, OLS, Elastic-net, or polynomial regression, should also be performed. Larger data sets with more complexity and higher data dimensions could also be used in the experiments.

\noindent\textbf{Support vector machine (SVM)} is one of the most popular supervised learning models. Koh et al.~\cite{koh2020understanding} studied poisoning attacks on SVM models with an RBF kernel on a dog vs. fish image classification dataset extracted from ImageNet. They construct adversarial versions of training images to maximize the loss on the test set. Similar to the previous attack on LR, this attack is also developed on top of the bi-level optimization problem, however focusing on the highest test loss as determined by an influence function. They performed experiments on the same data set and showed good results: with only 2 perturbed training images for each test image, the model produces incorrect predictions on $77\%$ of the 591 test images. However, the experiments were only performed on the state-of-the-art Inception v3 network but not on the SVM-RBF network. As indicated by Goodfellow et. al, RBF networks are shown to make the model more robust by adding non-linearity~\cite{goodfellow2015explaining}. One of the potential improvements could be to evaluate the attacks against a model with an added RBF layer to verify whether non-linearity provides further protections. Another worth mentioning drawback is that it requires access to the testing set, which seems to be an impractical assumption. The algorithm also runs on $O(np)$ where $n$ is the training set size and $p$ is the number of model parameters. An obvious conclusion is that this limits the attack ability on big data with big models. 

Another similar poisoning attack against SVM tries to craft poisoning samples that maximize the hinge loss incurred on a set of validation samples~\cite{ref7}. It focuses on indiscriminate attack against model integrity using the similar bi-level optimization technique. However, as it optimizes based on validation loss, it requires less knowledge of the model, with the cost of having a less effective attack compared to the other SVM attack.

\subsection{Deep Learning}
As deep neural networks (DNN) become more and more popular, recent studies have been studying on poisoning attacks on more complex machine learning models. There are many papers focusing on vision models. For instance, Yang et al.~\cite{yang2017generative} adapt the gradient-based technique to deep neural networks (DNN). They also proposed to leverage the Generative Adversarial Network (GAN) technique: they used an auto-encoder as the generator to generate poisoned data using a reward function based on the loss function, and the target NN model as a discriminator to receive the crafted data and calculate the loss with respect to the legitimate data. They reported competitive results on MNIST and CIFAR-10. However, their attack requires a white-box setting, which may not be realistic in practice.

There are also poisoning attacks targeting DNNs' model availability. Shumailov et al.~\cite{sponge} propose an attack that generates sponge examples, which significantly increase the energy consumption during model training. This leads to a longer run time during model inference, which negatively affects the model availability. Specifically, for NLP models, the attack aims to increase the computation dimension, causing an increase in the number of tokens needed to be processed; for computer vision, they aim to decrease data sparsity by exploiting the ReLU activation function, which will minimize the benefit of GPUs. The poisoning samples are created via a genetic masking technique, in which two parent samples are crossed over with a random mask. The attack is demonstrated to be transferable across hardware platforms and model architectures. They evaluate the sponge examples under white-box and black-box setting. They observe larger impact on NLP models than vision models. 

One may argue that the above attack targets the inference stage instead of the training phase, therefore it should not be considered as a poisoning attack. However, such poisoning samples could also be fed into a machine learning model during training instead, which disrupts the model and significantly slows down the training progress. The threat model of this attack may sound impractical if we move towards an attack during model training, however we believe this is a feasible attack against federated learning. 

Another attack on model availability is the reordering attack~\cite{reordering}. It is a very special paper because it focuses on availability at training time with clean data and labels. That is, no modification on any parts of the training samples, not just the labels. This attack leverages the fact that the training result of stochastic learning and batching is data order dependent. The authors introduce various ordering methods in which training samples are sorted based on the loss magnitude. By evaluating on both NLP and vision tasks using DNN models, the attack is shown to be effective by decreasing the test accuracy by a maximum of 64\%. 

The above two studies are very distinct from the majority of existing work on clean-label attacks: they are not built upon the bi-level optimization problem to craft poisoning samples that maximize some specific model loss, instead, they focus on several existing issues in machine learning training. The reordering attack aligns with the property of catastrophic forgetting, which is the tendency of a learning model to abruptly forget previously learned information when learning about new features. The sponge examples target the fact that GPUs have the advantage to quickly process sparse data in compared to CPUs. These literature open up a new direction for researchers to develop adversarial actions by turning an existing machine learning property (or even an advantage) into a vulnerability to be exploited.

\section{Conclusion and Comments}

In this section, we first summarize the existing works we study in this survey. We then provide some overall comments on the field of ML poisoning attacks. Lastly, we highlight some intriguing research questions as future directions.

\subsection{Summary}

\begin{table*}[]
\begin{tabular}{|l|l|l|}
\hline
      \textbf{Attack Model}              & \textbf{Advantages}                                                                                                                                                    & \textbf{Disadvantages}                                                                                                                                                                                                  \\ \hline
Labeling flipping   & \begin{tabular}[c]{@{}l@{}}- Guaranteed effectiveness\\ - Computational inexpensive\end{tabular}                                                              & \begin{tabular}[c]{@{}l@{}}- Requires access to sample labeling process\\ - May be easily detected by human examiners \\ or automated detectors\end{tabular}                                                   \\ \hline
Watermarking        & \begin{tabular}[c]{@{}l@{}}- Enable backdoor attacks to gain internal access\\ - Possible to perform model/data stealing\end{tabular}                         & \begin{tabular}[c]{@{}l@{}}- Requires fine-tuning on the hyperparameters \\ to achieve desired results\\ - May be easily detected by human examiners \\ or automated detectors\end{tabular}                    \\ \hline
Clean-label attacks & \begin{tabular}[c]{@{}l@{}}- Generalized well for various ML models\\ - Formalize the attack into solvable \\ mathematical optimization problems\end{tabular} & \begin{tabular}[c]{@{}l@{}}- The bi-level optimization problems are intractable\\ - Even with estimation and relaxed assumptions, \\ the derived objective functions are still expensive to solve\end{tabular} \\ \hline
\end{tabular}
\caption{A summary of the three major attack methodology on poisoning attacks against supervised machine learning models}
\label{table:attack_compare}
\end{table*}

Table~\ref{table:attack_compare} summarizes the advantages and disadvantages of various poisoning attack methodology on supervised machine learning. In general, labeling flipping models are relatively inexpensive, whereas the clean-label attacks are the most computational expensive method. However, labeling flipping create poisoning samples that are easily noticeable, which can be detected and discarded by human examiners or automated detectors. Here we observe a trade-off relationship between computational cost and the attack secrecy. On the other hand, we can see that clean-label attacks are more generalized and systematic, whereas labeling flipping and watermarking depend on heuristics or parameter tuning in order to adjust to produce the most effective poisoning samples. We also want to highlight the major challenge of clean-label attacks: many attacks in this category formalize their optimization problems into the bi-level objective functions, yet the exact solutions are intractable to be found. Therefore, many existing works use different approaches to provide estimations or to reduce the computation cost when searching for an influential poisoning sample. 

\subsection{Subtle Differences}

Referring back to the comparison table (Table~\ref{table:compare_table}), readers may observe some similarities among the existing works. For instance, the two labeling flipping attacks (Barreno et al.~\cite{ref2} and Paudice et al.~\cite{ref3}), the poison frog attack that uses watermarking in non-finetuning models only (Shafahi et al.~\cite{poison_frog}), the clean-label attack on SVM (Biggio et al.~\cite{ref7}), and the logirtisc regression attack (Jagielski et al.~\cite{jagielski2021manipulating}) all formalize their poisoning attacks into an optimization problem. Despite being seemingly similar, all studies presented in this survey have their unique and novel contributions towards the field of ML poisoning attacks. Here are some subtle differences that readers may not notice:

\begin{enumerate}
  \item The two labeling flipping attacks make changes in the sample labels, whereas the other attacks only modify training samples.
  \item The poison frog attack builds upon a training-target relationship and aims to find poisoning samples that ``collide'' with the target instance. It uses a forward-backward-splitting iterative procedure to solve the optimization problem. In fact, the optimization equation in this attack is not bi-level.  
  \item The attack targeting logistic regression requires additional steps in compared to other attacks. This is because other attacks are formalized as a classification problem while LR requires numerical predictions. Related work on classification models often randomly initialize a poisoning sample, and then proceed to update this sample by solving the optimization problem iteratively. In the LR paper, in fact, the authors leverage labeling flipping in the initialization phase. Therefore, we could say that the LR attack involves both labeling flipping and clean-label attack formalized as a bi-level optimization problem.
  
\end{enumerate}

\subsection{``Clean Labels'' vs. ``Clean Samples''}

In this section, we want to discuss about the definition of a ``Clean-label attack''. We want to highlight the fact that a \textit{clean label} is not necessarily a \textit{clean sample}. For instance, the poisoning samples generated in the aforementioned clean-label attacks all have clean labels, but only the re-ordering attack produces ABSOLUTE clean samples that are unmodified. 

\begin{figure*}
    \centering
    \includegraphics[scale=0.4]
    {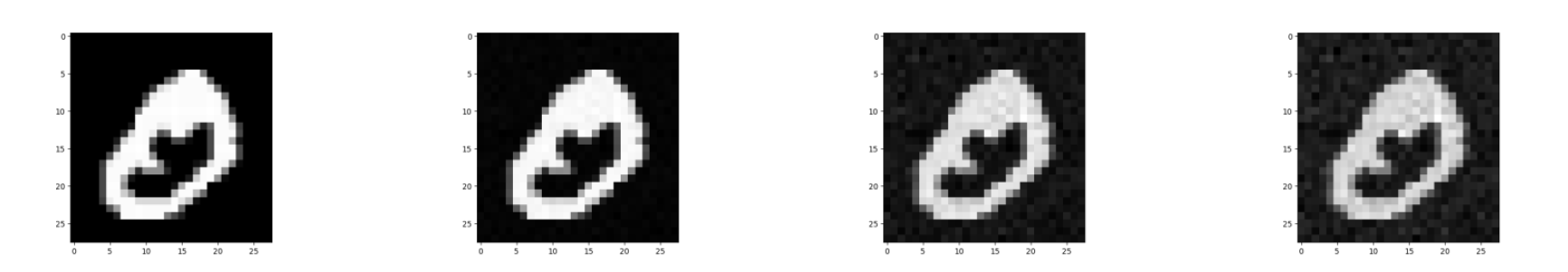}
    \caption{Example of a poisoned MNIST sample: the original digit, the poisoned digits with a poisoning iteration of $120$, $1200$, and $12000$, from left to right respectively. Although they should all be labeled as ``0'', human can observe obvious differences as the perturbation increases.}
    \label{fig:mnist_noise}
\end{figure*}

Intuitively, we may think of a clean poisoning sample as samples that are perturbed with adversarial noises yet remaining undetected if presented to a human labeler, for example, the adversarial examples generated by the Fast Gradient Signed Method (FGSM) attack~\cite{goodfellow2015explaining}. However, what about perturbations that are visible, but do not incur a change by the human labelers? As an example, Figure~\ref{fig:mnist_noise} shows four images of the same origin: the first one is the original image from MNIST, while the remaining ones are created by introducing poisoning perturbations into the original training data. It is obvious that there are noises added as the poisoning iteration increases. However, even with a very large number of noises, we as human labelers still recognize the label as it is originally. 

The challenge is defining a clean sample is to develop an appropriate metrics to evaluate how ``clean'' the sample is. To our knowledge, there is no such solid definition of a clean sample. Here we propose a new definition of ``clean samples''. A sample is \textit{clean} if and only if its creation process satisfies the following three conditions: 

\begin{enumerate}
  \item It does not require any control over the labeling of training data;
  \item It does not produce poisoning samples that are labeled differently by human labelers. That is, $y_s = y_h$ where $y_s$ is the sample label of $x$, and  $y_h$ is the label of the given sample $x$ by human.
  \item It does not involves specific feature information captured from a target instance. 
  
\end{enumerate}

As an example, a poisoning sample generated from the watermarking attack (with low opacity) may satisfy the first and second requirement, it does not meet the last requirement because it directly extract the features from a specific target instance. On the other hand, the re-ordering attack is a clean-sample attack, in which the training data are unchanged. 

\subsection{Future Work}

We summarize several research questions as future directions:
\begin{enumerate}
  \item What will be the ultimate solution for the bi-level optimization problem? Is there any possibility to reduce computational costs while crafting an effective poisoning sample?
  \item What are some potential metrics to evaluate how good the crafted samples are? Is this related to machine learning interpretability? What is the most appropriate metric to measure the negative impact of an individual poisoning sample on the learning model?
  \item How to quantify the ability of a given attack to ``hide'' from automated detection or human examiners? A related reference on PCA poisoning attack by Rubinstein et al.~\cite{2009_antidote} suggests that the ``boiled frog'' attack, in which small amounts of adversarial data are inserted in a gradual increasing trend in each iteration, is shown to prevent detection. Is it possible to integrate such mechanism to all existing work we described above?
  \item What are some machine learning properties that could be turned into potential vulnerabilities to be exploited? For instance, is it possible to poison an ML model by directing it to an exploding gradient or vanishing gradient? Is it possible to find a poor initial weights so that the models could hardly recover? Is it possible to move the learning weights towards a local minimal so it is stuck there with no learning progress?  
  \item As mentioned, poisoning attacks are especially harmful towards federated learning, where training samples are stored locally and the learning model updates as it trains on distributed client machines. It is also vulnerable to backdoor attacks, which can be easily achieved by leveraging the watermarking attack model. As known, one of the crucial characteristics of federated learning is that it aims to protect user data privacy. Does this protection on privacy come with a cost of being exploited by potential poisoning attack? What are some solutions to address this piratical challenge?  
  \item The ``poison frog'' attack could enable ``invisible'' backdoor attacks by simply uploading poisoning samples to openly available online platforms and waiting for these data to be scraped and used as training samples to big data ML models. This is in fact a serious challenge. As an example, the famous BERT language model is trained on Wikipedia, which can be easily modified by unauthorized online users with no trace. It is also impractical to examine the massive amount of online data by human experts. The only solution is to develop some automated detection systems to identify and discard such poisoning samples, but how?
  
\end{enumerate}

As a relatively new field, the research on ML poisoning attacks are still developing. We hope this survey paper provide some useful insights and guide readers to solve existing challenges in this area. In addition to poisoning attacks, there are also many other attacks towards ML systems, for instance, evasion attacks, privacy attacks, model stealing, adversarial examples in testing time, and so on. Researchers are also focusing on defenses against these attacks. For instance, a PCA-based defense against poisoning attack, proposed by Rubinstein et al.~\cite{2009_antidote}.  

In conclusion, this survey paper presents a comprehensive study on poisoning attacks against various supervised ML models. We analyze and compare the three major attack techniques, and highlight their advantages and disadvantages accordingly. We provide a brief discussion to identify current challenges in ML poisoning attacks, and provide guides for future directions. We believe that these studies are helpful for the machine learning and security community.

\section{Acknowledgements}
We thank Baochun Li for helpful feedback on this manuscript.

\newpage

\bibliographystyle{IEEEtran}
\bibliography{a}


% \end{thebibliography}


\end{document}